\documentclass[10pt,a4paper]{article} 
\def\vec#1{\mbox{\boldmath $#1$}} 
\usepackage{epsfig}
\begin{document} 
\textwidth=135mm 
 \textheight=200mm 
\begin{center} 
{\bfseries Quantum Condensates in Nuclear Matter: Problems} 

\vskip 5mm 
 
G. R\"opke, D. Zablocki 
 
\vskip 5mm 
 
{\small {\it Inst. Physik, Universit\"at Rostock, Rostock, Germany}} 
\\

\end{center} 
 
\vskip 5mm

\centerline{\bf Abstract}  
In connection with the contribution ``Quantum Condensates in Nuclear 
Matter'' some problems are given to become more familiar with the 
techniques of many-particle physics. 
%
\section{Bogoliubov transformation for superfluid state} 
The entropy operator $S$, which is related to the (grand 
canonical) statistical operator via $\rho = \exp [-S/k_B]$, has in 
second quantization the form  
$S = S_0 + S_1 + S_2 \dots$ where $S_0$ is a normalization constant, 
the single-particle contribution has the general form 
$$ 
S_1 = \sum_{ij}s(i,j) a^\dagger_i a_j + \sum_{ij}d(i,j) a_i 
a_j +\sum_{ij}d^*(i,j) a^\dagger_j a\dagger_i 
$$ 
where the Lagrange multipliers $s(i,j),d(i,j), d^*(i,j)$ are 
determined by the given averages when maximizing the entropy, the 
two-particle contribution and higher terms are neglected in the 
mean-field approximation considered here.
Assuming that $S_1$ is hermitean and that the normal term $s(i,j)$ is 
diagonalized, we have 
$$ 
S_1 = \sum_p s(p) a^\dagger_p a_p + \sum_p d(p,\bar p) a_p 
a_{\bar p} +\sum_p d^*(p) a^\dagger_{\bar p} a^\dagger_p 
$$ 
if we furthermore assume that the condensate is formed with a given 
spin and momentum state for the two-particle system ($p$ denotes 
momentum and spin quantum number).
In particular we can take $s(p) = \vec p^2/2m_\sigma - \mu_\sigma$, 
and the pair amplitude couples $p = \vec p,\sigma$ with the state 
$\bar p = -\vec p,-\sigma$  
\subsection*{Problem 1:} 
Find the transformation which diagonalizes $S_1$. 
\subsection*{Solution:} 
The Bogoliubov transformation for the creation and annihilation operators reads 
\begin{eqnarray} 
a_p = u_p b_p - v_p b_{\bar{p}}^\dagger  
&&
a_p^\dagger = u_p^* b_p^\dagger - v_p^*b_{\bar{p}}~, \\ 
a_{\bar{p}} = u_p b_{\bar{p}} + v_p b_p^\dagger  
&&
a_{\bar{p}}^\dagger = u_p^* b_{\bar{p}}^\dagger + v_p^* b_p~, 
\end{eqnarray} 
where the new operators obey the anticommutator relations 
\begin{eqnarray} 
\{b_p,b_{p'}^\dagger\} = \delta_{p,p'} &, 
& \{b_p,b_{p'}\} = \{b_p^\dagger,b_{p'}^\dagger\} = 0  
\label{com} 
\end{eqnarray} 
In order to get a canonical transformation, i.e. the anticommutator remains  
unchanged, we have to claim $\{a_p,a_{p'}^\dagger\}=\delta_{p,p'}$. 
%
%
\begin{eqnarray*} 
\{ a_p,a_{p'}^\dagger \}
&=&  
        u_p b_p u_{p'}^*b_{p'}^\dagger 
        -u_p b_p v_{p'}^*b_{\bar{p}'} 
        -v_{p} b_{\bar{p}}^\dagger u_{p'}^* b_{p'}^\dagger 
        +v_{p} b_{\bar{p}}^\dagger v_{p'}^* b_{\bar{p}'} \\ 
&&
        + u_{p'}^* b_{p'}^\dagger u_p b_p 
        -v_{p'}^* b_{\bar{p}'} u_p b_p
        -u_{p'}^* b_{p'}^\dagger v_p b_{\bar{p}}^\dagger 
        + v_{p'}^* b_{\bar{p}'} v_p b_{\bar{p}}^\dagger \\ 
&\stackrel{(\ref{com})}{=}&
(|u_p|^2+|v_p|^2)\delta_{p,p'} 
\end{eqnarray*} 
So it follows  
\begin{equation} 
|u_p|^2+|v_p|^2=1 ~. 
\label{test1} 
\end{equation} 
We now list the prefactors in $S_1$ where we make use of the  
anticommutator relations (\ref{com}). 
\begin{table}[h!] 
\begin{tabular}{ccc|cccc} 
&&
& $b_{\bar{p}}^\dagger b_{\bar{p}}$ & $b_p b_{\bar{p}}$ & $b_{\bar{p}}^\dagger b_p^\dagger$ & $b_p b_p^\dagger$  \\\hline 
$s(p)$ & : & 
$a_{\bar{p}}^\dagger a_{\bar{p}}^{}$ & $-|v_p|^2$ & $-|u_p|^2$ & $-u_p v_p^*$ & $-v_p u_p^*$ \\ 
$d(p)$ & : & 
$a_p a_{\bar{p}}$ & $u_p v_p$ & $-u_p v_p$ & $u_p^2$ & $-v_p^2$  \\ 
$d^*(p)$ & : & 
$a_{\bar{p}}^\dagger a_p^\dagger$ & $u_p^* v_p^*$ & $-u_p^* v_p^*$ & $-v_p^{*2}$ & $u_p^{*2}$ \\ 
$s(p)$ & : & 
$a_p a_p^\dagger$ & $|u_p|^2$ & $|v_p|^2$ & $-u_p v_p^*$ & $-u_p^* v_p$ 
\end{tabular} 
\caption{Coefficient matrix of the canonical (Bogoliubov) transformation} 
\label{coeff} 
\end{table} 
In order to achieve a diagonal transformation the inner columns of Table  
\ref{coeff} have to vanish, 
i.e. after substituting $s(t)=\epsilon_p$ and $d(p)=\Delta_p$ we have 
\begin{equation} 
-2\epsilon_p(u_p v_p^* + u_p^* v_p) 
+\Delta_p(u_p^2 - v_p^2) 
+\Delta_p^*(u_p^{*2} - v_p^{*2}) 
= 0  
\label{test2} 
\end{equation} 
One can easily check that 
\begin{eqnarray} 
u_p
&=&
\frac{1}{\sqrt{2}}\sqrt{1 + \frac{\epsilon_p}{\sqrt{\epsilon_p^2 + \Delta_p^2}}}~, \\  
v_p
&=&
\frac{1}{\sqrt{2}}\sqrt{1 - \frac{\epsilon_p}{\sqrt{\epsilon_p^2 + \Delta_p^2}}} 
\end{eqnarray} 
solve the equations (\ref{test1}) and (\ref{test2}) with $\epsilon_p$, 
$\Delta_{p}$ being real functions.
Furthermore we obtain 
$$ 
S_1(p)
=
\sqrt{\epsilon^2_p + \Delta_p^2}
\left( b_p^\dagger b_p + b_{\bar{p}}^\dagger b_{\bar{p}} - 1\right)~, 
$$ 
which is now diagonalized. 
\section{Transition from BCS to BEC} 
 
\subsection*{Problem 2:} 
Give the form of the wave function for the two-particle wave function 
including Pauli blocking for the Yamaguchi interaction (isospin 
singlet) at $E = 2\mu$. 
How changes the wave function in coordinate-space 
representation if we cross over from low densities (deuteron) to high 
densities (Cooper pairs)? 
\subsection*{Solution:} 
\begin{figure}[htb]
\epsfig{figure=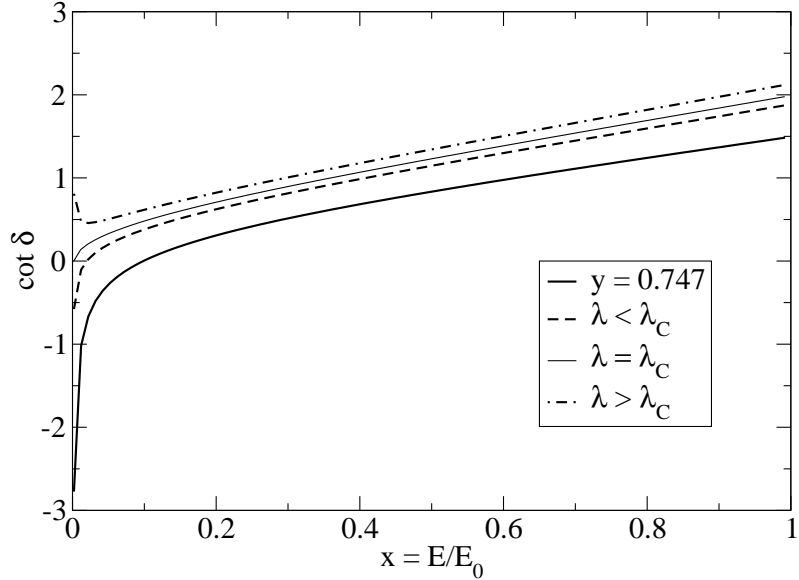,width=0.7\textwidth,angle=-90}
\caption{The scattering phase shift as a function of energy for several values of the coupling. $\lambda_C$ is defined in the text.}
\label{scattering}
\end{figure}

The Schr\"odinger equation for the two-particle problem including Pauli  
blocking reads 
\begin{equation} 
[E(1)+E(2)-E_{nP}]\psi_{nP}(12) 
+\sum_{1'2'}[1-f(1')-f(2')]V(12,1'2')\psi_{nP}(1'2')~=0, 
\label{2parti} 
\end{equation} 
where $E(1)= E(p_1) = p_1^2/2m_1$ denotes the kinetic energy of the 
single-particle  state $\{1\}=\{{\bf p_1},\sigma_1,\tau_1\}$  
abbreviating linear momentum, spin and isospin, respectively.
The Fermi distribution function 
$f(1)=\{\exp\{[E(p_1)-\mu_1]/T\}+1\}^{-1}$  
is characterized by the temperature $T$ and 
the chemical potential $\mu_1$. 
\\ 
The Yamaguchi potential is 
$$ 
V(12,1'2')
=
-\frac{\lambda_\sigma}{M} W((p_1 - p_2)/2)W((p_{1'} - p_{2'})/2) 
\delta_{p_1 + p_2,p_{1'} + p_{2'}} 
$$ 
with the formfactor $W(p) = (p^2 + \gamma^2)^{-1}$,  
the effective range $\gamma = 285.8484~ {\rm MeV}$ and the effective coupling  
constant $\lambda_{S=1,T=0} = 0.4144~{\rm fm^{-3}}$  
for the spin triplet, isospin singlet channel and $M$ is taken as the average nucleon mass.\\ 
After restriction to a two-nucleon system at rest in an isospin symmetric  
medium ($\mu_1 = \mu_2 = \mu$; $m_1 = m_2 = m$), 
the wave function satisfying the Schr\"odinger equation (\ref{2parti}) reads 
\begin{equation} 
\label{wf}
\psi_{\sigma}(p)= 
c_{\sigma}(T,\mu)~\frac{W(p)}{p^2/m - E_{\sigma}(T,\mu)}~ 
\end{equation} 
with the normalization factor $c_\sigma(T,\mu)$. 
Inserting (\ref{wf}) into (\ref{2parti}) one obtains the implicit
equation for the binding energy $E_\sigma(T,\mu)$
\begin{equation} 
\label{bse}
1 = \frac{\lambda_\sigma}{m}\sum_{p'}
\frac{W^2(p')[1-2f(p')]}{p'^2/m - E_{\sigma}(T,\mu)}
\label{E0} ~.
\end{equation} 
It is evident that the Pauli blocking factor in the kernel of (\ref{bse})
generates the $T$- and $\mu$- dependence of the binding energy.\\
The wave function in coordinate space is obtained by Fourier transformation,
\begin{equation} 
\psi_\sigma(r) 
=
\frac{1}{2\pi^2 r}\int_0^\infty{dp~p}~\psi_\sigma(p)\sin (p~r),
\end{equation}
with the final result
\begin{equation} 
\psi_\sigma(r) = \frac{c_\sigma~m}{4\pi ~(\gamma^2 + m E_\sigma)}
~\frac{e^{-\sqrt{- m E_\sigma}r} - e^{-\gamma r}}{r}~. 
\end{equation} 
A zero binding energy defines the transition from a bound state to a 
scattering state (Cooper pair). The corresponding transition from negative to 
positive energy eigenvalues entails a character change in the wave function 
to oscillatory behaviour.\\
The Schr\"odinger equation for the scattering problem (with the same restrictions as above) reads \cite{yama}
\begin{equation}
(k^2-p^2+i\epsilon)\psi(p)
=
-\lambda_\sigma W(p)\sum_{p'}[1-2f(p')]W(p')\psi(p')~, 
\label{2scat} 
\end{equation}
with the solution
\begin{equation} 
\psi(p) 
=
\delta(\mathbf{p-k}) - \frac{k^2 + \gamma^2}{2\pi^2}
\cdot \mathcal{F}_\sigma(k)
\cdot\frac{1}{\gamma^2+p^2}
\cdot\frac{1}{k^2-p^2+i\epsilon}
~ \label{solscat}
\end{equation} 
where $\mathcal{F}_\sigma(k)$ is the scattering amplitude and can be obtained by inserting (\ref{solscat}) into (\ref{2scat}). When we neglect Pauli-Blocking, $\mathcal{F}_\sigma(k)$ takes the form
\begin{equation}
\mathcal{F}_\sigma(k)
=
\left( -ik+\left( -\gamma+\frac{\gamma^2+k^2}{2\gamma}+\frac{(\gamma^2+k^2)^2}{2\pi^2\lambda_\sigma} \right) \right)^{-1}.\label{scat}
\end{equation}
Including Pauli-Blocking, the numerical solution is given in \cite{sch_roep}.\\ 
Equivalently, one can reprsent the solution in terms of the phase shift 
$\delta_\sigma(\mu,T)$, given by
\begin{equation}
\mathcal{F}_\sigma(k) 
=
\frac{e^{i\delta_\sigma}\sin\delta_\sigma}{k} =\frac{1}{-ik+k\cot\delta_\sigma}.
\end{equation}
Using (\ref{scat}) and $E=k^2/m$, one obtains the explicit form for the scattering 
phase shift $\delta_\sigma(E)$ from
\begin{equation}
\cot\delta_\sigma(E)
=
-\frac{1}{2\sqrt{x}}\left( 1 - x - y(1 + x)^2 \right)
\label{pijump}
\end{equation}
where we have introduced dimensionless variables $y=\gamma^3/(\pi^2\lambda_\sigma)$
and $x=E_\sigma/E_0$ with $E_0=\gamma^2/m$.\\
In order to discuss qualitatively the Mott-Effect of vanishing bound states we consider (\ref{2scat}) without Pauli-Blocking but with variable potential strength.
According to the Levinson theorem, at the critical coupling strength for the dissolution of the bound state, the scattering phase shift has to jump by $\pi$ at the threshold. 
This critical coupling strength can be calculated by solving (\ref{E0}) or (\ref{pijump}) respectively.\\
Let us consider the case when $E$ tends to zero ($x\to 0$).
(\ref{pijump}) gives
\begin{equation}
\lim_{x\to0}\cot\delta_\sigma=\frac{1}{2\sqrt{x}}(1-y)\mid_{x\to0}~.
\end{equation}
Using (\ref{E0}) and setting the binding energy to zero, we obtain the critical coupling $\lambda_c=\gamma^3/\pi^2$ which corresponds to $y_c=1$.
When approaching this critical coupling from both sides, we get
\begin{equation}
\lim_{\lambda\to\lambda_c\pm0}\cot\delta_\sigma(E) =  \pm\infty
\end{equation}
which corresponds to a jump of the scattering phase shift from $\pi$ to 0, in accordance with the Levinson theorem (illustrated in Figure \ref{scattering}).

We now want to consider the Pauli-Blocking energy shift by the example of the Deuteron.

%
\section{Pauli blocking for Gaussian bound states}    
\subsection*{Problem 3:} 
Construct a wave function for the bound states of 2, 3, 4 nucleons by 
minimizing the energy (Yamaguchi interaction) with respect to Gaussian 
wave functions!  
Calculate the Pauli blocking energy shift within first order 
perturbation theory!  
What is its dependence on temperature? 
\subsection*{Solution:} 
The four-particle Schr\"odinger equation reads 
\begin{eqnarray*} 
E_{nP}\psi_{nP}(1234) 
&=&
\left[E(1) + E(2) + E(3) + E(4)\right]\psi_{nP}(1234)\\ 
&& +\sum_{1'2'3'4'}\Big\{[1-f(1)-f(2)]V(121'2')\delta_{33'}\delta_{44'}\\ 
&& +[1-f(1)-f(3)]V(131'3')\delta_{22'}\delta_{44'}\\ 
&& +{\rm permutations}\Big\}\psi_{nP}(1'2'3'4')~. 
\end{eqnarray*} 
\subsection{$\alpha$-like clusters}
We make a Gaussian ansatz for the wave function, i.e.
\begin{equation}
\psi(r_1,r_2,r_3,r_4)
=
N\chi(R)e^{-\frac{1}{\alpha^2}((r_1-R)^2+(r_2-R)^2+(r_3-R)^2+(r_4-R)^2)},
\end{equation}
where $R=(r_1+r_2+r_3+r_4)/4$, or in Jacobi-coordinates
\begin{equation}
\psi(\xi_1,\xi_2,\xi_3,R)
=
N\chi(R)e^{-\frac{1}{\alpha^2}(\frac{1}{2}\xi_1^2+\frac{2}{3}\xi_2^2+\frac{3}{4}\xi_3^2)},
\end{equation}
where we have used the transformation rules
\begin{eqnarray}
R_k = \frac{1}{k}\sum_{i=1}^kr_i~,&&
\sum_{i=1}^{n}(r_i-R_n)^2 = \sum_{i=1}^{n-1}\frac{i}{i+1}\xi_i^2 ~.
\end{eqnarray}
The Fourier transformation of the wave function reads
\begin{eqnarray}
\psi(k_1,k_2,k_3,k_4) 
&=&
\int \psi(\xi_1,\xi_2,\xi_3,R)e^{-i(\xi_1q_1+\xi_2q_2+\xi_3q_3+KR)}d^3\xi_1d^3\xi_2d^3\xi_3d^3R\nonumber\\
&=&
N\chi(K)e^{-\frac{\alpha^2}{4}(\frac{2}{1}q_1^2+\frac{3}{2}q_2^2+\frac{4}{3}q_3^2)}\delta_{K,k_1+k_2+k_3+k_4}
\end{eqnarray}
where $K=k_1+k_2+k_3+k_4$ and $q_i$ are the conjugate momenta to $\xi_i$.
\subsection{Normalization}
The normalization factor can be determined as usual by calculating the expectation value 
of the square of the wave function:
\begin{eqnarray*}
\frac{1}{N^2} 
&=& 
\left[\frac{\Omega_0}{(2\pi)^3}\right]^3 \int e^{-\frac{\alpha^2}{2}(\frac{2}{1}q_1^2+\frac{3}{2}q_2^2+\frac{4}{3}q_3^2)}d^3q_1d^3q_2d^3q_3\\
&=&  
\left[\frac{\Omega_0}{(2\pi)^3} \int e^{-\frac{\alpha^2}{2}(\frac{2}{1}q_1^2+\frac{3}{2}q_2^2+\frac{4}{3}q_3^2)}dq_1dq_2dq_3 \right]^3\\
&=&  
\left[\frac{\Omega_0}{(2\pi)^3}  \sqrt{\frac{2}{\alpha^2}}\sqrt{\frac{1}{2}\frac{2}{3}\frac{3}{4}} 
\left[  \int_{-\infty}^\infty e^{-x^2}dx\right]^3\right]^3\\
&=&  
\left[\frac{\Omega_0}{(2\pi)^3}\right]^3\frac{\sqrt{2}^3\sqrt{\pi}^9}{\alpha^9}~,
\end{eqnarray*}
so it results
\begin{equation}
N^2=N_\alpha^2 = \left[\frac{(2\pi)^3}{\Omega_0}\right]^3\frac{\alpha^9}{\sqrt{2}^3\sqrt{\pi}^9}~.
\end{equation}

\subsection{Kinetic energy term}
\begin{eqnarray*}
\langle E_{kin} \rangle &=& N^2\sum_{k_1,k_2,k_3,k_4}\frac{1}{2m}(k_1^2+k_2^2+k_3^2+k_4^2)\mid\psi(k_1,k_2,k_3,k_4)\mid^2\\
&=& \sum_K\frac{1}{8m}\mid\chi(K)\mid^2K^2\\
&& +N^2\sum_{q_1,q_2,q_3}\frac{1}{2m}\left(\frac{2}{1}q_1^2+\frac{3}{2}q_2^2+\frac{4}{3}q_3^2\right)e^{-\frac{\alpha^2}{2}(\frac{2}{1}q_1^2+\frac{3}{2}q_2^2+\frac{4}{3}q_3^2)}\\
&=& E_{kin}^K + N^2\sum_{q_1,q_2,q_3}\frac{1}{2m}\frac{\partial}{\partial(-\frac{1}{2}\alpha^2)}e^{-\frac{\alpha^2}{2}(\frac{2}{1}q_1^2+\frac{3}{2}q_2^2+\frac{4}{3}q_3^2)}\\
&=& E_{kin}^K + N^2\frac{1}{2m}\frac{-1}{\alpha}\frac{\partial}{\partial\alpha}\frac{1}{N^2}\\
&=& E_{kin}^K + \frac{1}{2m}\frac{9}{\alpha^2}
\end{eqnarray*}
where we explicitly separated the motion of the total momentum $K$.
\subsection{Potential energy term}
For a separable potential we can calculate the potential part of the energy as
\begin{eqnarray*}
\langle E_{pot} \rangle
&=&
 -3\frac{\lambda_s+\lambda_t}{\Omega_0}\sum_{q_1,q_2,q_3,K,q_1^\prime} w(q_1)w(q_1^\prime)\psi(q_1,q_2,q_3,K)\psi(q_1^\prime,q_2,q_3,K)\\
&=&
 -3\frac{\lambda_s+\lambda_t}{\Omega_0}\sum_{q_1,q_2,q_3,K,q_1^\prime} w(q_1)w(q_1^\prime)N^2\mid\chi(K)\mid^2  \\\nonumber
 &&\times
 e^{-\frac{\alpha^2}{4}(\frac{2}{1}q_1^2+\frac{3}{2}q_2^2+\frac{4}{3}q_3^2)}e^{-\frac{\alpha^2}{4}(\frac{2}{1}{q_1^\prime}^2+\frac{3}{2}q_2^2+\frac{4}{3}q_3^2)}\\
&=&
 -3\frac{\lambda_s+\lambda_t}{\Omega_0}\frac{1}{e^{-\frac{\alpha^2}{2}2q^2}}\sum_{q,q^\prime}w(q)w(q^\prime)e^{-\frac{\alpha^2}{2}(q^2+{q^\prime}^2)}\\
&=&
 -3\frac{\lambda_s+\lambda_t}{\Omega_0}\frac{1}{e^{-\frac{\alpha^2}{2}2q^2}}\left[\sum_q w(q)e^{-\frac{\alpha^2}{2}q^2}\right]^2~.
\end{eqnarray*}
As an example we assume the potential to be of Gaussian type  $w(q) = \exp\{-q^2/\gamma^2\}$, this leads to
\begin{eqnarray*}
\langle E_{pot} \rangle &=& -3\frac{\lambda_s+\lambda_t}{(2\pi)^3}\frac{1}{e^{-\frac{\alpha^2}{2}2q^2}}\left[\int e^{-\frac{1}{\gamma^2}q^2}e^{-\frac{\alpha^2}{2}q^2}d^3q\right]^2\\
&=& -24\frac{\lambda_s+\lambda_t}{(2\pi)^3}\frac{\alpha^3\sqrt{\pi}^3}{\left(\frac{2}{\gamma^2}+\alpha^2\right)^3}~.
\end{eqnarray*}
So finally the total energy is given by
\begin{equation}
\langle H_{int}\rangle = \langle H\rangle-E_{kin}^K = \frac{9}{2}\frac{1}{m\alpha^2} -24\frac{\lambda_s+\lambda_t}{(2\pi)^3}\frac{\alpha^3\sqrt{\pi}^3}{\left(\frac{2}{\gamma^2}+\alpha^2\right)^3}
\end{equation}

\subsection{Model calculation}
\flushleft
We can now minimize this expression with respect to the variational parameter
$\alpha$ and fit the binding energy for the $\alpha$ particle by adjusting our model
parameters.
We obtain
\begin{eqnarray*}
\lambda_t = 1317.8 ~\rm MeV~ fm^{-3}~,&&
\lambda_s = 667 ~\rm MeV~fm^{-3}
\end{eqnarray*}
and therefore
\begin{eqnarray*}
\alpha = 8.288\cdot10^{-3}~ \rm MeV^{-2}~,
\end{eqnarray*}

which gives us a binding energy of the $\alpha$ particle of $-28.2$ MeV.

For the other bound states with less nucleons, this works in a similar way.
\subsection{Pauli blocking shift}
Now we want to calculate the Pauli blocking shift due to finite temperature.
We will consider a bound state of 2 nucleons.
For 3 and 4 nucleons this works similar but is more cumbersome.
From (\ref{E0}) we know that the exact wave function in the vacuum satisfies
\begin{equation} 
1 = \frac{\lambda_\sigma}{m}\sum_{p'}
\frac{W^2(p')}{p'^2/m - E^0_\sigma}
\end{equation}
In perturbation theory we expand the binding energy around its vacuum
value, i.e. $|E_\sigma| = E_\sigma^0 - \Delta E^{\rm Pauli}(T, \mu)$ 
where the minus sign is because
the binding energy is negative ($E_\sigma = -|E_\sigma|$).
\begin{eqnarray*}
1 
&=&
 \frac{\lambda_\sigma}{m}\sum_{p'}
\frac{W^2(p')[1-2f(p')]}{p'^2/m + E^0_\sigma-\Delta E^{\rm Pauli}(T,\mu)}\\
&=&
 \frac{\lambda_\sigma}{m}\sum_{p'}
\frac{W^2(p')}{p'^2/m + E^0_{\sigma}}
\frac{1-2f(p')}{1-\frac{\Delta E^{\rm Pauli}(T,\mu)}{p'^2/m + E^0_\sigma}}\\
&=&
 \frac{\lambda_\sigma}{m}\sum_{p'}
\frac{W^2(p')}{p'^2/m + E^0_\sigma}
\left(
        1 - 2f(p') 
        + (1-2f(p'))\frac{\Delta E^{\rm Pauli}(T,\mu)}{p'^2/m + E^0_\sigma}
        + \mathcal{O}\left((\Delta E^{\rm Pauli})^2\right)
\right)
\end{eqnarray*}
Dropping higher order terms we arrive at
\begin{eqnarray}
\Delta E^{\rm Pauli}(T,\mu)
&=&
\frac{\int |\psi(p)|^2 (p^2/m + E^0_\sigma)2f(p)p^2dp}
       {\int |\psi(p)|^2~p^2 dp}~,
\end{eqnarray}
we have assumed $2f(p')$ to be small against $1$ and used the expression for
the wave function $\psi(p)$ given in Eq. (\ref{wf}).

\end{document}